\documentclass{article}

\usepackage{color}
\usepackage{graphicx}             
\usepackage{bm}                      
\DeclareGraphicsExtensions{.jpg,.pdf}              
\usepackage{amsmath,amssymb}                                    

\begin{document}

\title{Biofilm mechanics and patterns}
\author{A. Carpio (Universidad Complutense de Madrid, Spain), \\
E. Cebri\'an (Universidad de Burgos, Spain), \\
D.R. Espeso (Centro Nacional de Biotecnolog\'{\i}a, CSIC, Spain), \\
P. Vidal (Universidad Complutense de Madrid, Spain)}

%
%
\date{July 17, 2017}

\maketitle

{\bf Abstract.} From multicellular tissues to bacterial colonies, three dimensional cellular structures arise through the interaction of cellular activities and mechanical forces. Simple bacterial communities provide model systems for analyzing such interaction. Biofilms are bacterial aggregates attached to wet surfaces and encased in a self-produced polymeric matrix. Biofilms in flows form filamentary structures that contrast with the wrinkled layers observed on air/solid interfaces. We are able to reproduce both types of shapes through elastic rod and plate models that incorporate information from the biomass production and differentiations process, such as growth rates, growth tensors or inner stresses, as well as constraints imposed by the interaction with environment. A more precise study of biofilm dynamics requires tackling water absorption from its surroundings and fluid transport within the biological system. This process alters the material properties of the biofilm and the overall stresses. We analyze whether poroelastic approaches can provide a suitable combined description of fluid-like and solid-like biofilm behavior.

\vskip 1mm

\section{Biofilm shapes}
\label{sec:bioshapes}
Understanding how cellular systems evolve to adopt different
shapes is an intriguing question which has motivated many theories.
Here, we try to unravel this process in simple living beings:
bacterial communities called biofilms. Environmental
conditions seem to play a key role inducing changes. 
Biofilms growing in  flows often form filaments constrained
by the surrounding geometry.
They may cross the current in corner flows \cite{stonepnas}
or wrap around tube walls forming helices \cite{carpioscirep}.
Instead, biofilms spreading on semisolid agar surfaces exhibit 
different types of wrinkles \cite{chaimrs,wilkingmrs}. 

To understand the development of a biofilm one must take into account its nature. 
In a biofilm, bacteria are glued together and 
to a surface by a self-produced polymeric substance: the EPS matrix. Once the 
biofilm is formed, it can be seen as a  biomaterial whose properties are controlled 
by the cellular activity \cite{chaimrs,wilkingmrs}.  We discuss here how the material
properties of a biofilm influence its shape in different environments.

\section{Filamentary structures in flows}
\label{sec:flows}

When spreading in flows, biofilms elongate with the
current forming threads. The shape of the thread adapts
to geometrical constraints, seeking to minimize
adequate energies. Its time evolution until an equilibrium
shape is reached can be described by discrete rod
models. We tackle here two different experimental
frameworks: biofilms in networks of
cylindrical tubes and biofilms in corner flows.

\subsection{Helical biofilms}
\label{sec:helical}

\begin{figure}[!hbt] 
\centering
\includegraphics[height=2.5cm]{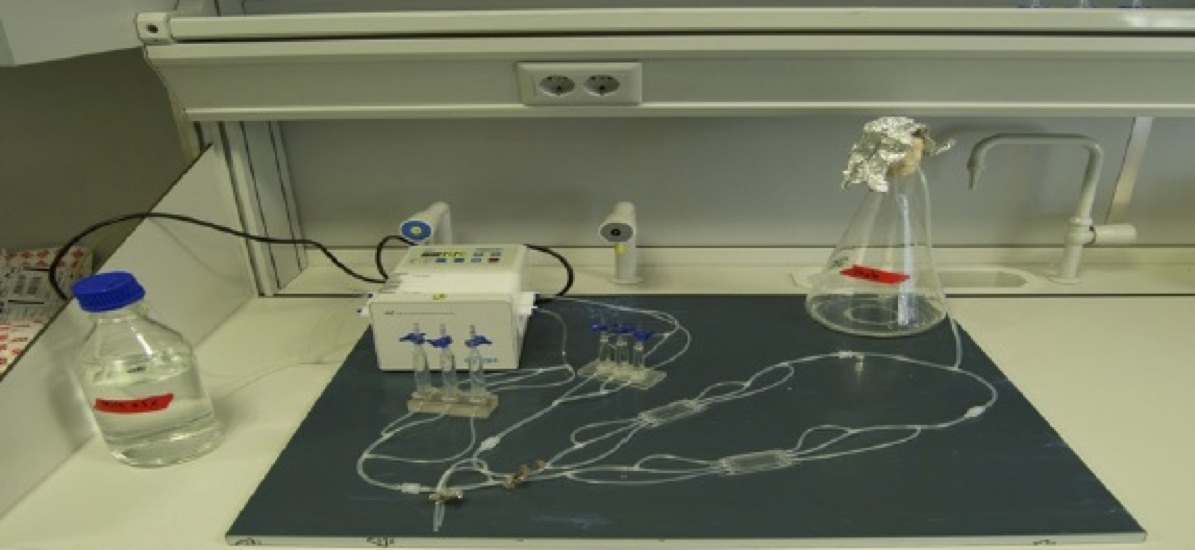}
\caption{Flow circuit. The fluid mixture flows from an initial reservoir to a pump (or a drip mechanism) that drives the liquid through a network of tubes, which may merge or split.}
\label{fig1}
\end{figure}

\begin{figure}[!hbt]
\centering
\includegraphics[height=2.5cm]{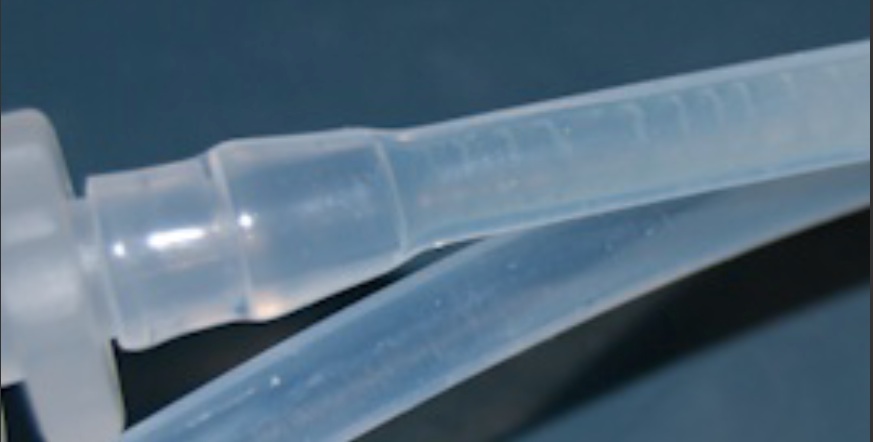}
\caption{Experimental image showing a biofilm helix wrapped
around the tube wall past the junction. No biofilm is formed
in the unperturbed branch. Bacteria were injected upstream the junction.}
\label{fig2}
\end{figure}

\begin{figure}[!hbt][b]
\centering
\includegraphics[angle=90,height=2.5cm]{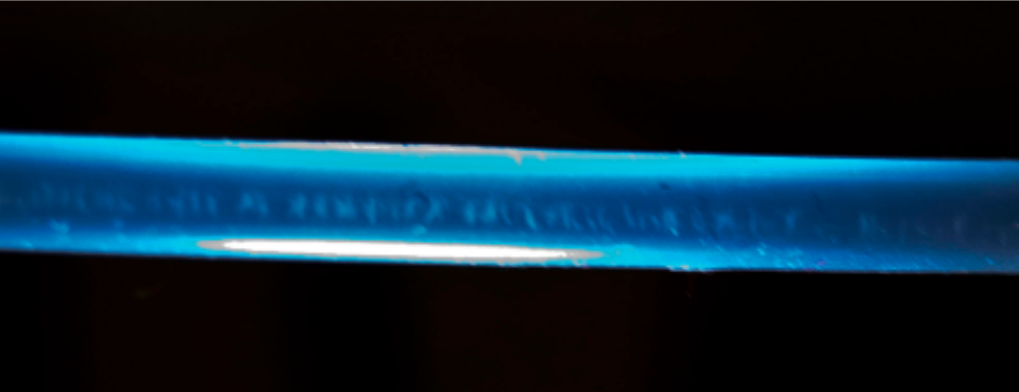}
\caption{Experimental image showing the onset of a helical instability along a biofilm 
thread in a 2mm diameter tube.}
\label{fig3}
\end{figure}

Consider the typical flow circuits used in medical systems, see Figure  \ref{fig1}. Injecting inside 
bacteria of the {\it Pseudomonas} genus, tubes fill with helical biofilms which wrap around the 
walls, see Figure \ref{fig2}. Even if the Reynolds number is fairly small ($Re \sim 1$), the presence of connectors and junctions produces diameter variations, locally narrowing the passage.
Vortices form past the stenosis. Vortical motion drive bacteria to the walls creating biofilm nucleation sites \cite{carpioscirep}. The biofilm then elongates following the streamlines until it undergoes a helical instability. Figure \ref{fig3} shows the initial stages of the helical instability. 

Elastic energies for filaments admit both straight and helical minimizers Ê\cite{helices}. The presence of constraints that forbid the straight equilibrium prompt the appearance of helical structures \cite{carpioscirep}. 
A coiling effect is usually active at some end due to the presence of constrictions. When the biofilm hits a surface, deceleration results in coiling. Also, the presence of vortices at the stenosis may induce a helical beating of the thread. 
Additionally, biomass production causes a continuous length increase of the thread between constrictions. Helical shapes allow to allocate the excess length. Uninterrupted biomass production  fosters the coarsening of the helical instability until it reaches the tube walls. More biomass is then allocated by narrowing  the helix pitch. Notice that a filament of length $L_f$
wraps around a tube of radius $r_t$ and length $L_t$ forming $k$ steps of pitch $L_t/k$
when $L_f^2 \sim L_t^2 + 4 \pi^2 r_t^2 k^2$.

\begin{figure}[!hbt]
\centering
\hskip 2.5cm (a) \hskip 5cm (b) \\
\includegraphics[width=5.8cm]{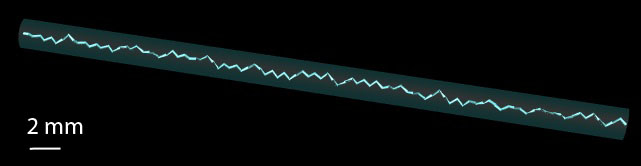}
\includegraphics[width=5.8cm]{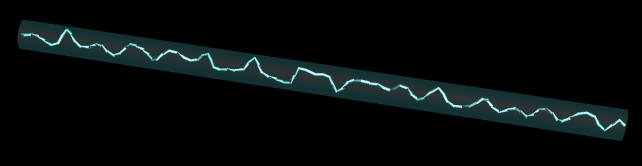} \\
\textcolor{white}{.} \hskip 2.4cm (c) \hskip 5cm (d) \\
\includegraphics[width=5.8cm]{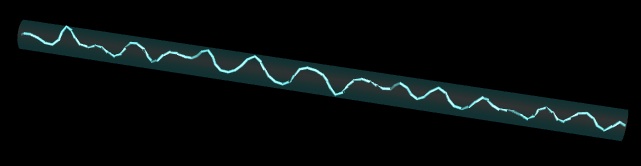}
\includegraphics[width=5.8cm]{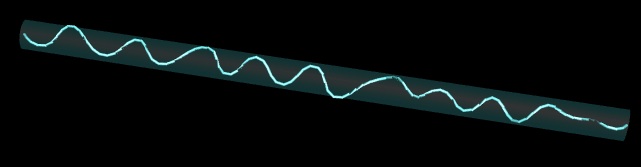} 
\caption{Snapshots illustrating the  in silico development and coarsening of a helical 
instability.}
\label{fig4}
\end{figure}

Figure \ref{fig4} illustrates this dynamic process.  A filament subject to twist develops a helical instability that coarsens until the helix reaches the tube wall thanks to a continuous length increase.  These simulations are performed using the discrete rod model described
in Section \ref{sec:discreterod} \cite{carpioscirep}. 

\subsection{Discrete rod framework}
\label{sec:discreterod}

\begin{figure}[!hbt]
\centering
\includegraphics[width=7.5cm]{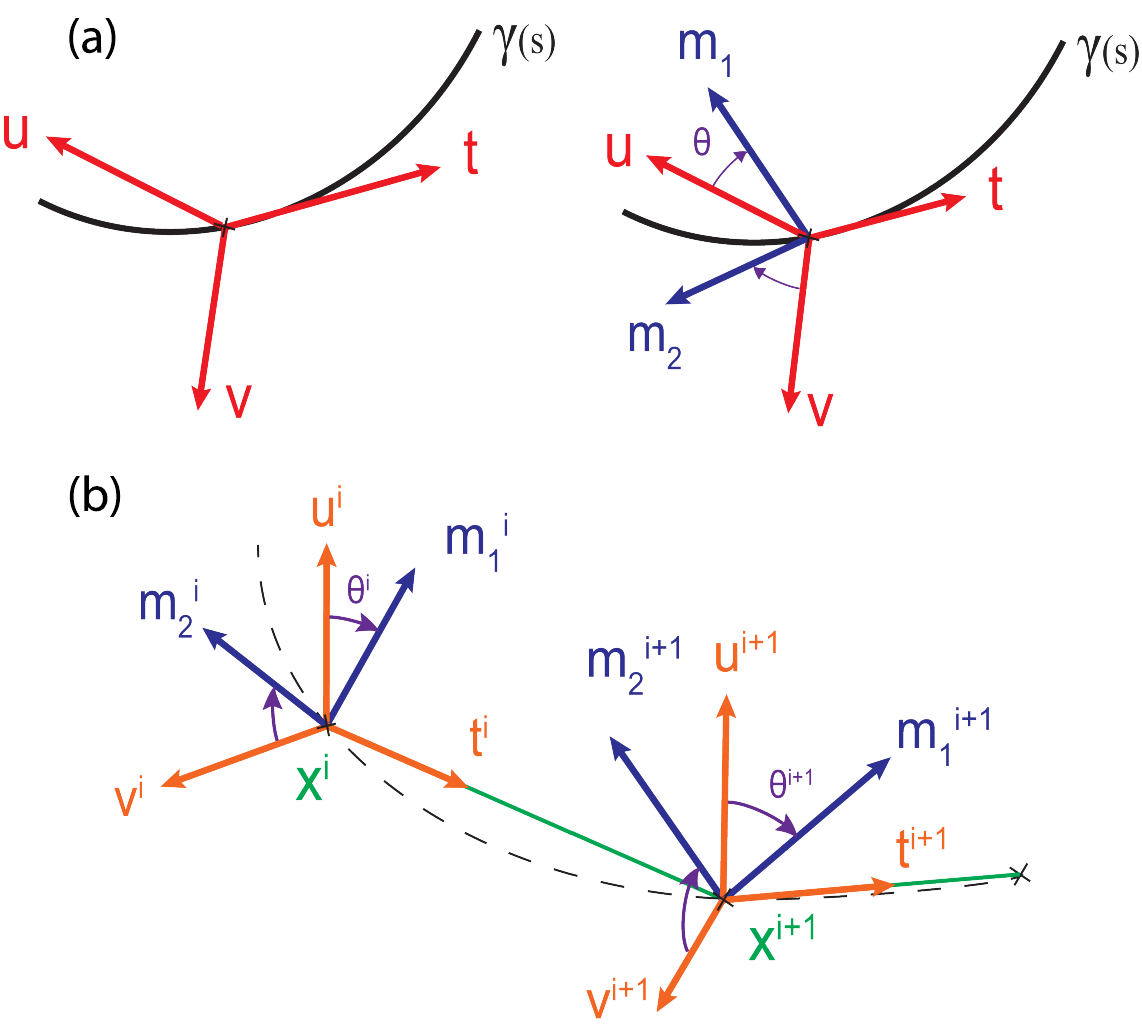}
\caption{(a) Continuous description of a filament \cite{langermaterial}.
The Bishop frame $\{{\mathbf t}, {\mathbf u}, {\mathbf v}\}$ 
defines the Òrest orientationÓ at any point of the filament
centerline $\gamma(s)$, parametrized by the arc length in $I\!\! R^3$.
Rotating $\{{\mathbf u}(s), {\mathbf v}(s)\}$ 
an angle $\theta(s)$ around ${\mathbf t}(s)$ we obtain
the material frame characterizing the local orientation 
$\{{\mathbf t}(s), {\mathbf m}_1(s), {\mathbf m}_2(s)\}$.
(b) Discrete description of a filament \cite{discreterods}.  The centerline is 
discretized as a set of points $\{{\mathbf x}^0, {\mathbf x}^1,\ldots, 
{\mathbf x}^{n+1}\}$ and segments ${\mathbf e}^i = 
{\mathbf x}^{i+1} - {\mathbf x}^i$. 
Setting ${\mathbf t}^i = {{\mathbf e}^i
\over \|{\mathbf e}^i\|}$, the unit tangent vector per edge,
a local orthonormal material frame 
$\{{\mathbf t}^i, {\mathbf m}_1^i, {\mathbf m}_2^i\}$
is assigned to each point. }
\label{fig5}
\end{figure}

A filament is a geometric shape whose length is much larger than the rest of its dimensions. Any deformation of its cross section is expected to be small compared with variations of the total length. This fact motivates the description of a biofilm thread as a unidimensional curve 
$\gamma$ (the centerline, which characterizes its position), plus a reference system  at each
point $\{ {\bf t}, {\bf m}_1, {\bf m}_2 \}$ (the material frame, which measures the twist).
With this description, the movement of the thread can be fully captured: stretching and bending are computed by deforming the centerline, whereas twisting is captured by the orientation of the material frame. 
For dynamic simulations we use a discrete rod model  \cite{langermaterial,discreterods}. The filament is discretized using a sequence of nodes ${\mathbf x}^i$, $i=0,\ldots,n+1$, along the curve $\gamma$, and a reference system at each one, see Figure \ref{fig5}. This frame is obtained at each location twisting the Bishop frame (a fixed untwisted frame) a certain angle 
$\theta^i$. The dynamics of the discrete filament is then governed by equations for the angles 
$\theta^i$, and for the node positions ${\mathbf x}^i$.  We detail the procedure next.

\subsubsection{Reference frames}
\label{sec:frames}

We assign  to each point a local orthonormal frame (the material frame) 
$\{ {\mathbf t}^i, {\mathbf m}^i_1, {\mathbf m}^i_2\}$, $i=0,\ldots,n$, 
describing the centerline orientation as follows. Let us denote by
${\mathbf e}^i= {\mathbf x}^{i+1}- {\mathbf x}^{i}$, $i=0,\ldots,n$, 
the straight segments joining the points
$\{ {\mathbf x}^0, {\mathbf x}^1,..., {\mathbf x}^{n+1} \}$.
The unit tangent vector associated to each edge is then
${\mathbf t}^i= { {\mathbf e}^i \over \| {\mathbf e}^i \|}$, where
$\| \|$ denotes the euclidean norm.
Assuming the Bishop frame $\{{\bf t}^i, {\bf u}^i, {\bf v}^i\}$ known, the
vectors ${\mathbf m}^i_1, {\mathbf m}^i_2$ are obtained rotating
${\bf u}^i, {\bf v}^i$ an angle $\theta^i$ in the plane orthogonal to
${\bf t}^i$:
\begin{eqnarray}
\mathbf{m_1^i} = \cos(\theta^i)   \mathbf{u}^i + \sin(\theta^i)   \mathbf{v}^i,  \quad
\mathbf{m_2^i} = - \sin(\theta^i)  \mathbf{u}^i + \cos(\theta^i)   \mathbf{v}^i. 
\label{materialdiscrete}
\end{eqnarray}

To define a Bishop frame we choose ${\bf u}^0 \perp {\bf t}^0$ and set  ${\bf v}^0= {\bf t}^0 \times {\bf u}^0.$ The frames at the remaining edges are constructed by parallel transport \cite{discreterods}. We set 
\begin{eqnarray}
{\bf u}^i = P_i ({\bf u}^{i-1}), \quad {\bf v}^i= {\bf t}^i \times {\bf u}^i,
\label{bishopdiscrete} 
\end{eqnarray}
where $P_i$ are rotation matrices about the curvature binormal defined by:
\[
P_i ({\bf t}^{i-1})= {\bf t}^{i}, \quad  P_i ({\bf t}^{i-1} \times {\bf t}^{i})= 
{\bf t}^{i-1} \times {\bf t}^{i}. 
\]
If ${\bf t}^{i-1} = {\bf t}^{i}$, $P_i$ is the identity. The condition 
${\bf u}^0 \perp {\bf t}^0$ must be maintained during the simulation.
This is guaranteed when ${\bf t}^0$ is clamped. Otherwise, it can 
be reestablished by parallel transport in time (instead of space).

\subsubsection{Equations for the angles}
\label{sec:angles} 

The equations for the angles follow from energy arguments.
When the undeformed configuration of the filament is straight and its elastic response is isotropic, the elastic energy due to torsion and bending takes the form \cite{discreterods}:
\begin{eqnarray} 
E= \sum_{i=1}^n \beta {(\theta^i - \theta^{i-1})^2 \over \overline{\ell}^i} +  \sum_{i=1}^n {\alpha \over 2 \overline{\ell}^i} \sum_{j=i-1}^i 
\| {\bf w}_i^j - \overline{\bf w}_i^j  \|^2,
\label{energydiscrete}
\end{eqnarray}
where $\alpha$ and $\beta$ are the bending and torsion moduli, respectively.
We may set $\alpha=E_YI$ and $\beta=JG$, being $E_Y$ the Young modulus of the thread, $I$ the second moment of area, $G$ the shear modulus of the thread and $J$ the torsional rigidity constant. If we consider a thread composed by an isotropic elastic material, the Young modulus and shear modulus are related by the Poisson coefficient $\nu$ as $G=\frac{E_Y}{2(1+\nu)}$. For a filled cylinder we have $J = I$, hence $\beta=\frac{E_Y}{2(1+\nu)}I$.

In formula (\ref{energydiscrete}), $\overline{\ell}^i$ is the length of the segments 
$\overline{\bf e}^i= \overline{x}^{i+1} - \overline{x}^{i}$
in a reference undeformed  configuration 
$\{ \overline{\mathbf x}^0, \overline{\mathbf x}^1,..., \overline{\mathbf x}^{n+1} \}$. 
The vectors ${\bf w}_i^j$, $\overline{\bf w}_i^j$, $j=i-1,i$, are 
material curvatures in the deformed and undeformed configurations, respectively:
\begin{eqnarray}
{\bf w}_i^j = \left( (\kappa{\bf b})_i \cdot {\mathbf m_2^j}, 
- (\kappa{\bf b})_i \cdot {\mathbf m_1^j} \right)^t,
\quad
(\kappa{\bf b})_i  = { 2 {\mathbf e}^{i-1} \times   {\mathbf e}^{i}
\over  \| \overline{\mathbf e}^{i-1}\| \|\overline{\mathbf e}^{i}\| + 
{\mathbf e}^{i-1} \cdot  {\mathbf e}^{i} },
\label{mcurvaturediscrete}
\end{eqnarray}
where $\kappa {\bf b}$ is the curvature binormal. For an undeformed straight
shape $\overline{\bf w}_i^j=0$.
The general form of the elastic energy for anisotropic rods that adopt a nonstraight 
undeformed shape is given in  \cite{discreterods}.

The material frame is updated in a quasistatic way. Imposing 
\begin{eqnarray}
{\partial E \over \partial \theta^i}=0,
\label{quasistaticdiscrete}
\end{eqnarray} 
for all segments $i$ not fixed by a boundary condition, this system of equations determines the angle configuration that minimizes the energy of the thread.  
Clamped ends are accounted for assigning the material frame for $i=0$, $i=n$. No boundary condition corresponds to a stress free end.

\subsubsection{Equations for the positions}
\label{sec:positions} 

We keep track of the filament position displacing the nodes according to Newton's second law:
\begin{eqnarray}
{\cal M} {d^2{\bf x} \over dt^2} = - {d E \over d {\bf x}} + {\bf F},
\label{newtondiscrete}
\end{eqnarray}
where $\mathbf{F}$ represents the external forces and 
$ - {d E \over d {\bf x}}$ the elastic forces. Explicit formulas for the
elastic forces are given in \cite{discreterods}.
${\cal M}$ denotes the $3(n+2) \times 3(n+2)$ mass matrix, we set 
${\cal M}=m {\cal I}$, where ${\cal I}$ is the identity matrix. $\mathbf x
= (\mathbf x^0, \ldots, \mathbf x^{n+1})$
denotes the ordered sequence of 3D node coordinates. Notice
that $\mathbf x^i= (x^i_1,x^i_2,x^i_3).$

To approximate the solution of system (\ref{newtondiscrete}), we apply a Verlet integrator 
to estimate the displacements and velocities, and then enforce an inextensibility constraint for each segment by using a manifold projection method \cite{inextensibility}. This constraint facilitates a stable evolution from a numerical point of view.

For each new time $t_{k+1}=t_k +h$, the Verlet scheme provides the prediction:
\begin{eqnarray}
\tilde{\bf v}_{k+1} = {\bf v}_{k} - h {\cal M}^{-1} {\bf F}({\bf x}_{k}), \quad
\tilde{\bf x}_{k+1} = {\bf x}_{k} + h \tilde{\bf v}_{k+1},
\label{verlet}
\end{eqnarray}
starting from previous values ${\bf x}_{k}, {\bf v}_{k}$, where
${\bf v}_{k}=\dot{\bf x}_k$ and $h$ is the time step. The projection method
works as follows. We set ${\bf y}_0=\tilde{\bf x}_{k+1}$. At each step $j$, 
we compute the next value ${\bf y}_{j+1}={\bf y}_{j}+ \delta {\bf y}_{j+1}$, where
\begin{eqnarray}
\delta {\bf y}_{j+1} = - h^2 {\cal M}^{-1} \nabla {\bf C}({\bf y}_j)^t \, \delta {\ell }_{j+1},
\label{multipliers}
\end{eqnarray}
and ${\bf C}({\bf y})$ defines the system of constraints.  The vector 
$\delta {\ell }_{j+1}$ solves the linear system
\begin{eqnarray}
h^2 \nabla {\bf C}({\bf y}_j) {\cal M}^{-1} \nabla {\bf C}({\bf y}_j)^t  \, \delta {\ell }_{j+1}
=  {\bf C}({\bf y}_j).
\label{multipliers2}
\end{eqnarray}
The iteration stops if
$\|{\bf C}({\bf y}_{j+1})\| \leq \varepsilon$ for the desired tolerance $\varepsilon >0$.
The constraint enforcing velocity and position are  then
\begin{eqnarray}
{\bf v}_{k+1}= {1\over h} ({\bf y}_{j+1}-\tilde{\bf x}_{k+1}), \quad
{\bf x}_{k+1} = {\bf x}_{k} + h  {\bf v}_{k+1}.
\label{enforcing}
\end{eqnarray}
If inextensibility is the only constraint, ${\bf C}$ is defined by the system of equations
$  \|{\bf e}^i\|^2 / \|\overline{\bf e}^i\|   - \|\overline{\bf e}^i\|=0$, for each edge $i$.

\subsubsection{Coupling to the flow}
\label{sec:flow}

The force exerted by a fluid undergoing a given undisturbed flow on a long slender body is analyzed in  \cite{cox}. 
The following asymptotic formula in terms of the ratio of the of the cross-sectional radius to the body length is given:
\begin{eqnarray}
\frac{{\bf f}^t}{2\pi} \!=\mu_f \lambda \int_0^1 \Big( \!\Big[ \frac{(\textbf{U} \!-\! \textbf{U}^*)^t}{\ln \, \kappa} \!+\! \frac{(\textbf{U} \!-\! \textbf{U}^*)^t \ln(2)}{(\ln \, \kappa)^2} \Big]  \cdot [ {\bf t} \, {\bf t}^t-2 {\cal I} ] \!+\! \nonumber \\
\frac{\frac{1}{2}(\textbf{U} \!-\! \textbf{U}^*)^t}{(\ln \, \kappa)^2} \cdot [ 3 {\bf t} \, {\bf t}^t \!-\! 2 
{\cal I} ] \Big) ds,
\label{force}
\end{eqnarray}
where \textbf{f} is the force acting on a filamentous body of length $\lambda$,  
$\textbf{U}$ and $\textbf{U}^*$ are the velocities of the unperturbed fluid and the thread, respectively, at the position $\gamma(s)$ and ${\cal I}$ is the $3\times 3$ identity matrix. 
We denote by $\mu_f$ the viscosity of the fluid, 
$\gamma(s)$ the position of the thread centerline, $s$  the arclength of the thread 
($0< s < 1$), ${\bf t}(s)=\frac{d \gamma(s)}{ds}$  the tangent vector at the position 
$\gamma(s)$ and  { $\kappa=\frac{r}{L}$ the ratio
between cross-sectional radius $r$ to characteristic thread length $L$}.

This relation allows us to directly calculate the fluid force ${\bf f}^i$ acting on each node of the thread by using the difference of velocities between the fluid and the thread, the tangent vector on each node and the aspect ratio of the thread, an idea already exploited for filaments in 2D
corner flows in \cite{stonefluids}. Notice that, at each node $\mathbf x^i$, the thread velocity 
$\mathbf U^*(\mathbf x^i,t)=\mathbf v^i(t)$.
In absence of other forces, ${\bf F}=({\bf f}^0,\ldots,{\bf f}^{n+1})$ in equations (\ref{newtondiscrete}).

Biofilm filaments live inside tubes of a certain shape. A simple way to incorporate this 
restriction is a penalty method. The idea is to include in the force term ${\bf F}$
in equations (\ref{newtondiscrete}) additional forces ${\bf f}_w$ supported on the 
tube walls that points inside the tube and acts on any node hitting the wall, sending it 
back inside. Alternatively,  we might set the position equal to the effective
maximum radius and reset the velocity equal to zero.

\subsubsection{Increasing the length}
\label{sec:length}

Increase in length of a filament can be due to the combined
effect of different mechanisms: biomass production, biomass adhesion, 
elastic elongation, swelling... To reproduce an
increase in length at a certain rate we enlarge the segments 
joining nodes in a controlled way, redefining the reference 
lengths at the same time. Directional mass addition may be
represented adding nodes at an edge and redefining the 
reference configuration each time a node is added. 

In practice, we alternate steps in which we solve the equations
for the evolution of the discrete rod with steps in which we increase the length 
of the edges or the number of nodes, and reset the reference configuration 
before computing again the evolution of the enlarged filament. 

\subsubsection{Overall procedure}
\label{sec:overall}

Summarizing, to compute the evolution of a rod we proceed in the following steps:
\begin{itemize}
\item Initialization:
\begin{itemize}
\item Define the Bishop frame at edge 0: $({\bf t^0}, {\bf u^0}, {\bf v^0})$.
\item Set the position of the undeformed centerline:  $\overline{\mathbf x}^0, 
\overline{\mathbf x}^1,..., \overline{\mathbf x}^{n+1} $.
\item Select the initial position and velocity of the centerline: $(\mathbf x^0, \dot{\mathbf x}^0), (\mathbf x^1, \dot{\mathbf x}^1), $ $..., 
(\mathbf x^{n+1}, \dot{\mathbf x}^{n+1})$.
\item Enforce the boundary conditions for the filament at the initial and final nodes.
\item Set the material curvatures using Eq. (\ref{mcurvaturediscrete}).
\item Set the material frame by means of Eqs. (\ref{bishopdiscrete}),
(\ref{materialdiscrete}) and (\ref{quasistaticdiscrete}).
\end{itemize}
\item Iteration for each new time step:
\begin{itemize}
\item Compute the elastic forces $- {d E \over d {\bf x}}$ acting on the centerline, and possible additional forces ${\bf F}$ (see section \ref{sec:flow}).
\item Integrate Newton equations for the centerline (\ref{newtondiscrete})
enforcing inextensibility and possible additional constraints (see section \ref{sec:positions}).
\item Update the Bishop frame using Eq. (\ref{bishopdiscrete}).
\item Update the quasistatic material frame by means of Eqs. (\ref{quasistaticdiscrete}) and (\ref{materialdiscrete}).
\item Eventually, increase the length of the edges or the number of nodes, and reset the reference configuration (see section \ref{sec:length}).
\end{itemize}
\end{itemize}

The simulations  shown in Fig. \ref{fig4} start from a straight filament placed at the center
of the tube. The initial node positions are randomly perturbed to ensure
a slight initial excess length with respect to the tube length. A twist angle is imposed at the
filament edges. The length of the filament is slowly increased as the helix 
develops to foster the coarsening process \cite{carpioscirep}. The unperturbed
fluid velocity profile obeys a radial Hagen-Poiseuille distribution.
However, the fluid force does not seem to play a role
in helix formation, which is driven by elastic forces. It only causes slow downstream 
motion of the whole structure. Therefore, we may set it equal to zero to simplify
the study of helix development. The force term 
${\bf F}$ would only account for the presence of the walls in this case, constraining
the helix radius. On the contrary, fluid forces are essential to produce the filaments crossing 
corner flows described in Section \ref{sec:corner}.

\subsubsection{Nondimensional equations}
\label{sec:dimensionless}

It is convenient to nondimesionalize  equations (\ref{energydiscrete}),(\ref{newtondiscrete}) 
and (\ref{force}) for numerical purposes. The change of variables  
$x = { \lambda} x'$ , $t = T t'$,  
${\bf U} = U_0 {\bf U}'$, $E =  {\alpha \over { \lambda}} E'$, 
${\bf  F} = \mu_f { \lambda} U_0  {\bf F}'$ yields:
\begin{eqnarray}
{d^2{\bf x'} \over dt'^2} = - {T^2 \over m { \lambda}^2} 
{d E \over d {\bf x'}} 
+ {T^2 \over m { \lambda}}  {\bf F} = - 
{\alpha T^2 \over m { \lambda}^3} {d E' \over d {\bf x'}} 
+ {\mu_f U_0 T^2 \over m}  {\bf F'},
\label{newtondiscretedimensionless}
\end{eqnarray}
where the force term includes the force exerted by the fluid ${\bf f}$ plus
possible penalty forces ${\bf f}_w$ at the walls, that is,
${\bf F}={\bf f} + {\bf f}_w$.
In view of the definition (\ref{energydiscrete}) of the energy $E$, this change 
introduces the controlling parameters  $\alpha' = {\alpha T^2 \over m { \lambda}^3}$,  
$\beta' = {\beta T^2 \over m { \lambda}^3}$ and $\delta'={\mu_f U_0 T^2 \over m}$,
$U_0$ being a characteristic velocity.
Determining ranges of values of $\alpha'$, $\beta'$ that lead to different types of filamentary structures we would obtain ranges for $\alpha$, $\beta$ whenever the 
density $\rho$ and radius $r$ of the filaments are experimentally quantified.
Working with one dimensional filaments, we neglect the cross-sections. A three-dimensional cylindrical thread of density $\rho$, radius $r$, and length {$ L >> r$}
is approximated in this setting by a discrete rod with $n+2$ nodes and $n+1$ edges, 
with mass { $m = \rho \pi  r^2  L /(n+1)$}.  From identities (\ref{energydiscrete}) and (\ref{newtondiscretedimensionless}), the characteristic time associated to the elastic deformation of the thread can be estimated as $T_{elast}=\sqrt{\frac{m{ \lambda}^3}{E_YI}}$. 
If this value is chosen as characteristic time in our system and we use formula 
(\ref{force}) we arrive at:
\begin{eqnarray}
{d^2{\bf x'} \over dt'^2} = -  {d E' \over d {\bf x'}} 
+ \eta{\bf F'},
\label{newtondiscretedimensionless2}
\end{eqnarray}
where $\eta=\frac{\mu_f U_0{ \lambda}}{\frac{E_YI}{{ \lambda}^2}}$  is the ratio of viscous forces to elastic forces, see \cite{stonefluids}.

\begin{figure}[!hbt]
\centering
(a) \hskip 5cm (b) \\
\includegraphics[height=5cm]{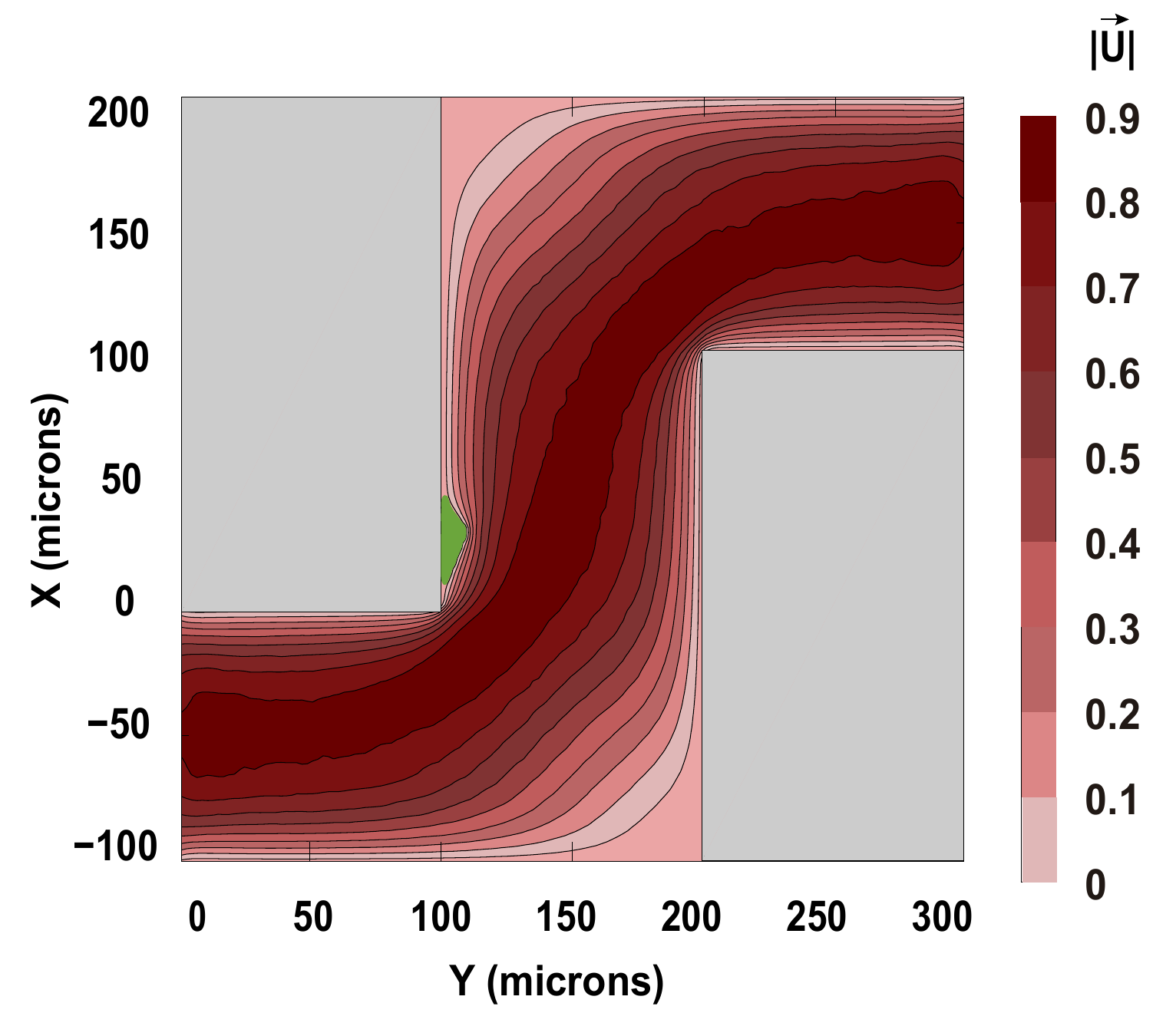} \hskip 5mm
\includegraphics[height=4.5cm]{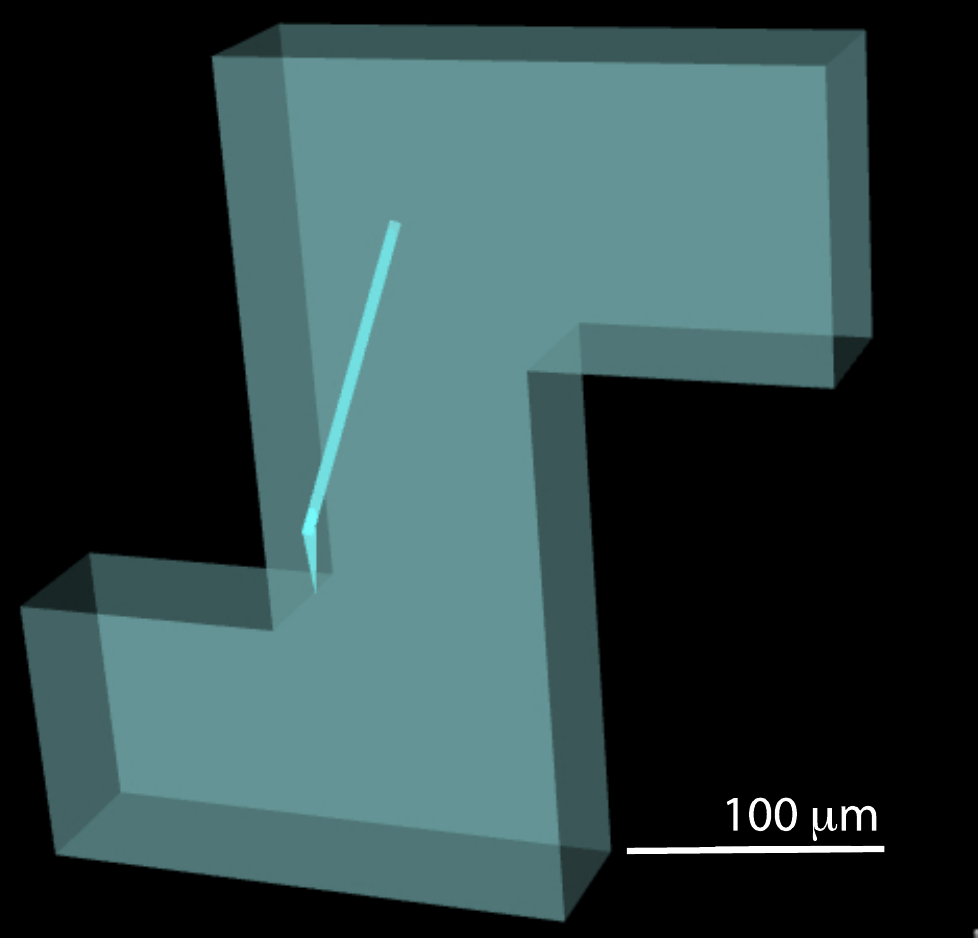}
\caption{
(a) Normalized velocity modulus in the mid-plane of the there dimensional channel.
(b) Initial filament configuration in the corner flow simulations.}
\label{fig6}
\end{figure}

\subsection{Biofilm threads in corner microflows}
\label{sec:corner}

\begin{figure}[!hbt]
\centering
(a) \hskip 5cm (b) \\
\includegraphics[width=5.5cm]{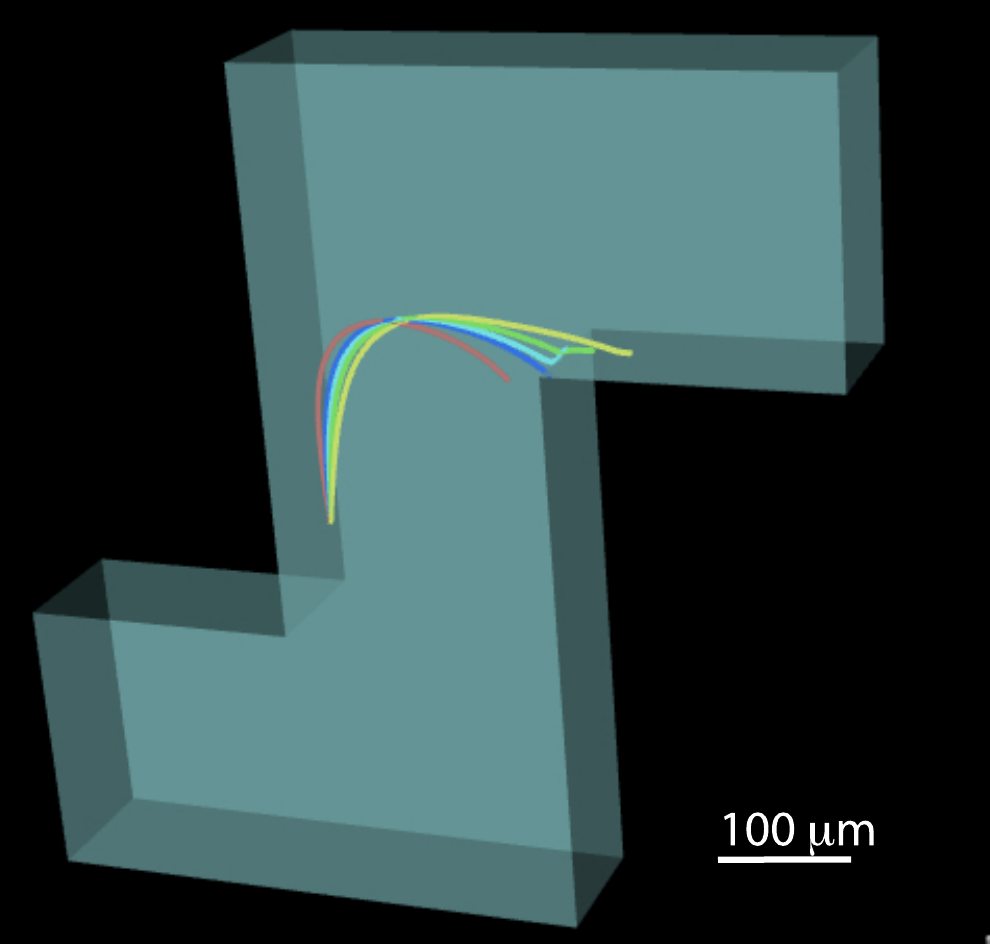}
\includegraphics[width=5.5cm]{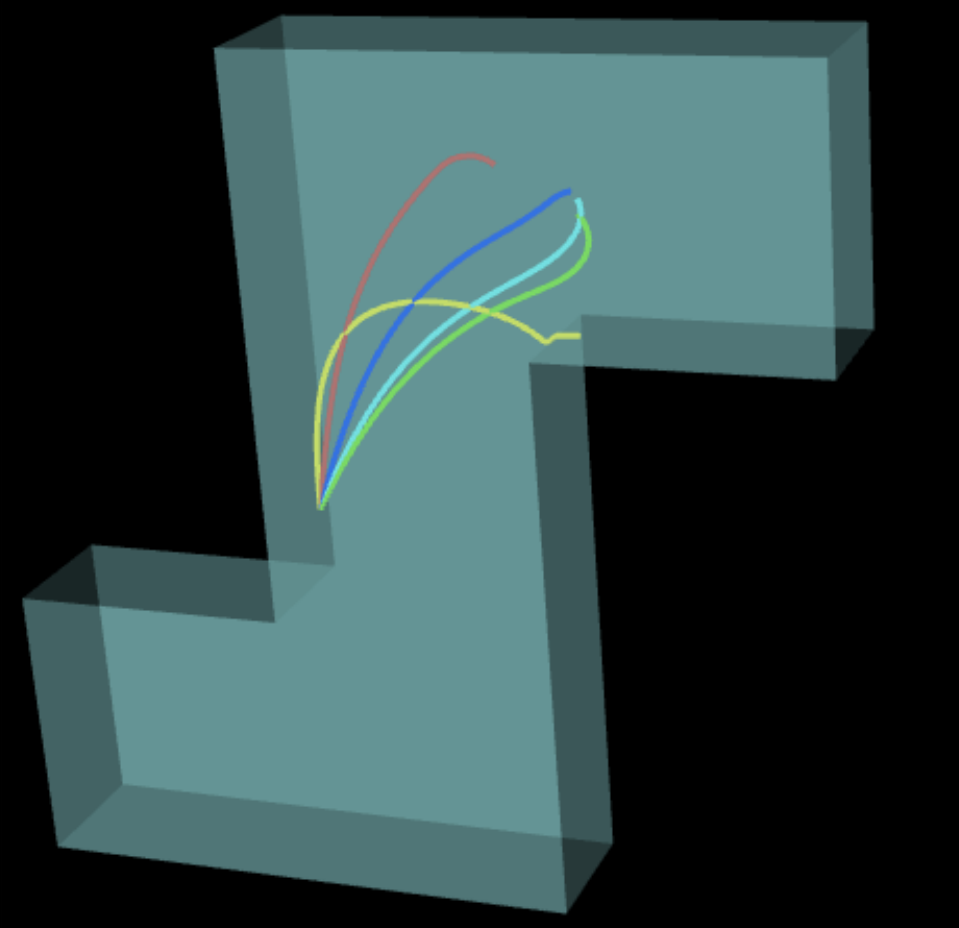}\\
\vskip 2mm
\includegraphics[width=5cm]{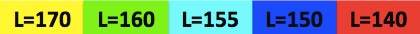}
\hskip 5mm
\includegraphics[width=5cm]{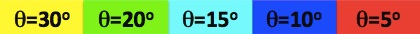}
\caption{Effect of the initial thread geometry on its dynamics during the simulation: 
(a) Length $L$ and  
(b) angle $\theta$. 
Parameter values are $\alpha'=1$, $\beta'=0.358$, $n=100$, $\eta=3000$, $\kappa=0.05$, $\Delta t'=10^{-3}$ time units.
A final state joining opposite corners is reached depending on the initial orientation and length, although these constraints may change with the initial velocity of the thread.}
\label{fig7}
\end{figure}
 
The situation described in Section \ref{sec:helical} for helical biofilms is partly reminiscent of 
the observations made with {\it Pseudomonas Aeruginosa} in a laminar corner microflow \cite{stonepnas}. 
Bacteria are driven  to the wall by small secondary vortices past the corner, creating nucleation sites. Once a biofilm seed forms, threads made of bacteria joined  by EPS matrix are issued.  
Initially, they align with the  streamlines, like streamers. 
Unlike the case of helical biofilms, which are triggered by elastic forces and constrained by 
the tube walls, biofilm threads cross the streamlines of corner flows driven by the fluid force
acting on them, as a result of the interaction fluid-structure. 
A two-dimensional model of an elastic filament in a corner flow shows that under certain conditions filaments cross the main stream and reach the opposite corner \cite{stonefluids}.
When the filament is long enough, it reaches the opposite corner adopting the equilibrium shape of an elastic rod in a corner flow. As in the case of  helices, the final configuration 
seems to be a minimum of an elastic energy.

The 3D model summarized in Section \ref{sec:discreterod} reproduces this behavior provided we include in the equations of motion (\ref{newtondiscrete}) the force due to the fluid \cite{cox}, given by (\ref{force}). 
The pressure driven fluid velocity field ${\mathbf U}$ may be computed using finite element software such as COMSOL multiphysics in the unperturbed channel geometry, see Fig. \ref{fig6}(a). We use this reference flow field during the whole simulation, ignoring perturbations due to the presence of the thin biofilm thread, as in \cite{stonefluids,cox}.   Fig. \ref{fig6}(b) represents the initial biofilm configuration used in the simulations.
Keeping the same parameters as in the experiments \cite{stonefluids}, we
can study the influence of variables like the initial angle or length on the filament evolution. 
Figure \ref{fig7} shows some possible configurations. To obtain it, we integrated the evolution equation (\ref{newtondiscretedimensionless2}) 
setting the characteristic spatial length $\lambda= 1 \mu$m and the characteristic time 
of the system equal to the elastic deformation time $T_{elas}$, which yields $\alpha'=1$, $\beta'=\beta/\alpha$, $\delta'=\eta$.

As observed in \cite{stonefluids}, the parameter $\eta$ regulates the effect on the fluid on the thread dynamics. For values of $\eta \leq 4000$, the contact of the thread with the opposite corner is only dependant of the filament initial position and length. A minimum length and angle are needed for the thread to reach the opposite corner.  Within the range $4000<\eta<8000$ the fluid can substantially modify the trajectory of the thread. For $\eta > 8000$ the fluid strongly drags the thread in the direction of the stream,  avoiding  contact with the opposite corner. A steady position for the thread parallel to the streamlines was found for $\eta \sim 30000$. These results are obtained setting the initial velocity of the thread equal to the fluid velocity at the node positions.  The initial thread velocity affects the results. Setting it equal
to zero, very low values of $\eta$ would be required to cross the streamlines.

If we start from a short thread and implement directional mass increase adding nodes at an edge and redefining the reference configuration each time a node is added, then we can see the filament grow as it moves towards the opposite corner.

\section{Wrinkled biofilms on agar}
\label{sec:wrinkled}

Whereas biofilms in flows tend to form filamentary structures,
biofilms spreading on agar/air interfaces adopt wrinkled shapes
\cite{asallypnas,chaimrs,wilkingmrs}. Descriptions of their
behavior may be made more precise than in the previous
case due to an increasing amount of experimental evidence.
Cell death has been shown to play a role on the onset of wrinkles
in {\it Bacilus Subtilis} biofilms. The biofilm is formed by bacteria immersed 
in a polymer matrix, which gives the mixture a certain elastic cohesion. 
Dividing cells produce compression stresses. In addition, cells may 
die due to biochemical stress associated with high cell density, high 
waste and toxin concentration, and lack of resources. 
Dead areas allow to relieve that stress forming wrinkles. 
This explains the onset of wrinkles \cite{asallypnas} but not the 
branching arrangements observed \cite{carpiopre}. These arrangements
can be understood incorporating information on cellular activity in
mechanical models of biofilm expansion on a substrate, as we
explain next.

\subsection{F\"oppl-Von Karman models}
\label{sec:vonkarman}

Let us consider a biofilm layer spreading on an agar substratum.
We can reproduce wrinkle branching in the expanding biofilm resorting to 
F\"oppl-Von Karman descriptions of the interface biofilm/agar \cite{benamar,huang,ni}: 
\begin{eqnarray}
{\partial \xi \over \partial t} &=& {1 - 2 \nu_v \over 2 (1-\nu_v)} {h_v \over \eta_v}
\Bigg[ D (- \Delta^2 \xi + \Delta C_M)  
 +  h {\partial\over \partial x_{\beta}} \left(  \sigma_{\alpha,\beta}({\bf u}) 
{\partial \xi \over \partial x_{\alpha}}\right) \Bigg]
-{\mu_v \over \eta_v} \xi, \label{plategrowth1bis} \\
{\partial {\bf u} \over \partial t} &=& {h_v h \over \eta_v} 
\nabla \cdot {\bm{\sigma}({\bf u}) } - {\mu_v \over \eta_v} {\bf u}, \label{plategrowth2bis}  
\end{eqnarray}
where $h_v$ is the thickness of the viscoelastic substratum and $\mu_v$, 
$\nu_v$, $\eta_v$  its rubbery modulus, Poisson ratio, and  viscosity, respectively. 
The bending stiffness is $D={Eh^3 \over 12 (1-\nu^2)}$, where
$E$ and $\nu$ represent  the Young are Poisson moduli of the biofilm, whereas
$h$ is the film thickness. In these equations, $\xi$ stands for the out of plane 
displacement and ${\bf u}$ the in plane displacement. $\alpha$ and $\beta$ stand
for $x,y$ and summation over repeated indices is intended. Stresses $\bm{\sigma}$ and 
strains $\bm{\varepsilon}$ are defined in terms of in-plane displacements  ${\bf u}=(u_x,u_y)$ \cite{benamar,mora}:
\begin{eqnarray}
 \varepsilon_{\alpha,\beta}={1\over 2} \left( {\partial u_{\alpha} \over \partial x_{\beta}} 
+ {\partial u_{\beta} \over \partial x_{\alpha}}  
+ {\partial \xi \over \partial x_{\alpha}} {\partial \xi \over \partial x_{\beta}}
\right) +  \varepsilon_{\alpha,\beta}^0, 
\label{strain}\\
 \sigma_{xx}= {E \over 1-\nu^2} (\varepsilon_{xx}+ \nu \varepsilon_{yy}), \quad
 \sigma_{xy}= {E \over 1+\nu} \varepsilon_{xy}, \quad
 \sigma_{yy}= {E \over 1-\nu^2} (\varepsilon_{yy}+ \nu \varepsilon_{xx}). 
\label{stress}  
\end{eqnarray}
The residual strains $\varepsilon_{\alpha,\beta}^0$ are expressed in terms 
of the growth tensor \cite{benamar} as:
\begin{eqnarray}
\varepsilon_{\alpha,\beta}^0= -{1\over 2} \left( {\cal G}_{\alpha \beta} + 
{\cal G}_{\beta \alpha} + {\cal G}_{z \alpha} {\cal G}_{z \beta} \right),
\label{residualstress}
\end{eqnarray}
and should be computed  from cellular activity in the spreading biofilm configuration.

\subsection{Bacterial activity}
\label{sec:bacterial}

Bacterial activity can be represented exploiting different agent based models.  Cellular automata descriptions, for instance, provide a simple framework allowing for an easy transfer 
of information into macroscopic models. The biofilm is divided in cubic tiles, each of them containing a few cells. To simplify further, we may identify each tile with one cell.
This approach has two advantages. First, we can use the same grid of tiles to discretize the equations for the relevant chemical concentrations and the displacements, and then solve them numerically.
Second, we can calculate the growth tensors due to cell division, death, and other processes, 
and use them to estimate the residual stresses that enter the F\"oppl-Von Karman
equations for the deformations. 
We have to decide for each cell which is its status. It may secrete chemicals, deactivate, 
divide creating a newborn cell that displaces the rest or die, being eventually reabsorbed by 
the rest. This may be done resorting to dynamic energy budget descriptions \cite{deb} or according to probabilities that depend on the relevant concentrations \cite{carpiopre}.

When there is an excess of oxygen, the concentration of nutrients $c_n$ becomes the limiting concentration that restricts biofilm growth. The evolution of the concentration $c_n$ in the
biofilm/agar system is governed by
\begin{eqnarray}
c_{n,t} - {\rm div} (D_{n}  \nabla c_n) = r_n(c_n), \label{cnb}
\end{eqnarray}
where $r_n(c_n) = - \tilde k_n {c_n\over c_n+K_n}$, and $\tilde k_n$ is the
uptake rate, equal to $k_n$ at each alive cell location and zero otherwise.
$D_n$ and $K_n$ denote the diffusion  and half-saturation coefficients. 
No flux boundary conditions are imposed at the interface with air.
The evolution of the concentration of waste $c_W$ in the
biofilm/agar system is governed by
\begin{eqnarray}
c_{w,t} - {\rm div} (D_{w}  \nabla c_w) = r_w(c_w), \label{cnb}
\end{eqnarray}
where $r_w(c_w) =  \tilde k_w $, and $\tilde k_w$ is the
waste production rate, equal to $k_w$ at each alive cell location and zero otherwise.
No flux boundary conditions are imposed at the interface with air. The diffusion
coefficient $D_w$ may vary across the biofilm/agar system.

In the simplest cellular automata approach, tiles  ${\cal C}$ occupied by alive cells are assumed to divide with probability \cite{hermanovic}:
\begin{eqnarray}
P_d({\cal C}) = {c_n({\cal C}) \over c_n({\cal C}) + a_n}, \label{pdiv}
\end{eqnarray}
$c_n$ being the limiting concentration and $a_n>0$. Newborn cells inside the biofilm are reallocated by pushing existing cells in the direction of minimum mechanical resistance, that is, the shortest distance to the biofilm-air interface. Taking  the concentration of waste $c_w$ at a location as an indicator of death, a cell ${\cal C}$ is scheduled to die with probability:
\begin{eqnarray}
P_w({\cal C}) = {c_w({\cal C}) \over c_w({\cal C}) + a_w}, \label{pdead}
\end{eqnarray}
$c_w$ being the waste concentration and $a_w>0$. Dead cells surrounded by enough 
alive neighbors may be reabsorbed by the rest, and its place occupied by a newborn cell. Otherwise, necrotic regions are created.
This process may be further refined to account for cell differentiation into producers of
different types of autoinducers \cite{carpiopre,deb}.

For a fixed distribution of cell types, the concentrations $c_n$ and $c_w$ relax
fast to stationary values, which may be approximated by explicit finite difference
schemes. Once the concentration values are calculated, we go through all the
cells forming the film, creating new cells or killing existing ones with the selected
probabilities. Then, a growth tensor may be defined at each tile by keeping track 
of all the new tiles created and the direction in which their predecessors where 
shifted. First we introduce a vector 
${\bf w}=(w_1,w_2,w_3) a,$ where $a$ is the tile size. $w_1$ is evaluated at each 
location by adding $\pm 1$ cumulatively  for each tile shifted in the $x$ direction in the 
positive or negative sense, respectively. $w_2$ and $w_3$ are calculated in a 
similar way, along the $y$ and $z$ directions, respectively. The resulting vector 
$\mathbf w$ is normalized to have norm $a$. Next, we compute $\nabla {\bf w}$ 
approximating the derivatives by finite differences. To estimate the growth tensor
${\cal G}(x,y)$ we average 
all the contributions from $\nabla {\bf w}(x,y,z)$ varying $z$.

\begin{figure}[!hbt]
\centering
\hskip 1cm (a)  \hskip 4.5cm (b) \\
\includegraphics[height=3cm]{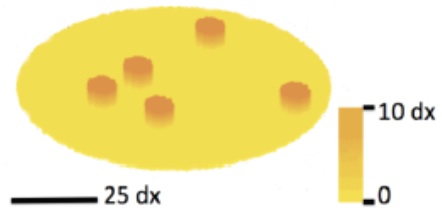}
\includegraphics[width=3cm,height=3cm]{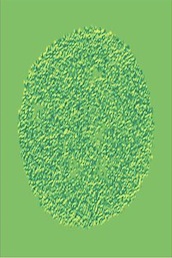}\\
 (c) \hskip 2.5cm (d) \hskip 2.5cm (e) \\
\includegraphics[width=3cm,height=3cm]{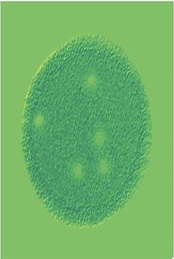}
\includegraphics[width=3cm,height=3cm]{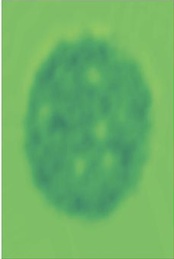}
\includegraphics[width=3cm,height=3cm]{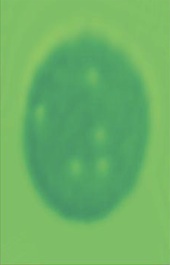}
\caption{(a) Biofilm containing  regions where the cell density is higher.
$dx$ represents the spatial step. 
(b) $\varepsilon_{xx}^{0,av}$ component  of the residual strain tensor due to growth
with $N=1$.
(c) Averaged $\varepsilon_{xx}^{0,av}$ component with $N=100$.
(d) Filtered $\varepsilon_{xx}^{0,fil}$ component with $N=1$.
(e) Filtered $\varepsilon_{xx}^{0,fil}$ component with $N=10$.
The  depressions correspond to the initial mounds and appear due to cell death caused by 
lack of resources. Strains are higher in the outer ring due to higher availability of resources, 
which results in higher division rates. The same scale of colors is used in all of them
ranging from $3$ (light yellow) to $-3$ (dark green).}
\label{fig8}
\end{figure}

\subsection{Residual strains}
\label{sec:residual}

The residual strains defined in (\ref{residualstress}) can be computed using the growth
tensor ${\cal G}$ introduced in Section \ref{sec:bacterial}. However, stochastic variations
make it unsuitable to be inserted directly in the F\"oppl-Von Karman equations (\ref{plategrowth1bis})-(\ref{plategrowth2bis}) because they cause numerical instability.

To smooth out the residual strains and visualize the underlying spatial variations, we
average them over a number of runs of the step in which new cells are created or
killed according to the selected probabilities, keeping the same initial configuration
in all of them: 
\begin{eqnarray}
\varepsilon^{0,av}= {1\over N}\sum_{j=1}^N \varepsilon^{0,j},
\end{eqnarray} 
where $\varepsilon^{0,j}$ stands for the residual strain at trial $j$.
Performing such ensemble averages  for $N$ large enough the averaged strains reproduce 
spatial variations reflecting cellular activity, see Figure \ref{fig8} (b)-(c). The resulting 
average becomes smoother as the number of runs $N$ increases. However, the 
computational cost of this process is high.

Instead, we  filter the residual fields using image processing techniques. This
strategy yields smooth approximations with a clear spatial structure averaging just a 
few runs, see Figure \ref{fig8}. The idea is to formulate a denoising problem: given
an observed field $f^{obs}=f+n$, we seek the underlying smooth structure $f$ obtained removing the noise $n$. To solve this problem we apply 
a split Bregman method to a ROF (Rudin, Osher, Fatemi) model of the denoising
problem \cite{torres}. The ROF model consists in solving the variational problem:
Find $f$ minimizing 
\[
\int |\nabla f| + {\mu \over 2} \int |f-f^{obs}|^2 = TV(f) + {\mu \over 2} \| f-f^{obs} \|_{L^2}^2,
\] for $\mu>0$
large. The split Bregman reformulation adds the constraint $d=\nabla f$, sets
$s(b,f,d)= \int |b + \nabla f -d|^2 $ and introduces the iteration:
\begin{eqnarray}
(f^{(k+1)},d^{(k+1)}) = {\rm Argmin}_{(f,d)} \{ |d|
+ {\mu \over 2} \| f^{obs} - f \|_{L_2}^2
+ {\lambda \over 2} s(b^{(k)},f,d)
\}, \nonumber \\
b^{(k+1)} = b^{(k)} + \nabla f^{(k+1)} - d^{(k+1)}. \nonumber
\end{eqnarray}
We split the minimization procedure to solve for each variable
separately:
\begin{eqnarray}
f^{(k+1)} = {\rm Argmin}_{f} \{ {\mu \over 2} \|f^{obs} - f \|_{L_2}^2
+ {\lambda \over 2} s(b^{(k)},f,d^{(k)})
\}, \nonumber \\
d^{(k+1)} = {\rm Argmin}_{d}  \{ |d|
+ {\lambda \over 2} s(b^{(k)},f^{(k+1)},d)\},
 \nonumber \\
b^{(k+1)} = b^{(k)} + \nabla f^{(k+1)} - d^{(k+1)}. \nonumber
\end{eqnarray}
The first functional is differentiable, therefore, we can write the Euler-Lagrange 
equation and evaluate $f^{(k+1)}$ with a Gauss-Seidel  method.
The second optimization problem can be solved using 
shrinkage operators:
\begin{eqnarray}
d^{(k+1)} = \mbox{\rm shrink} (b^{(k)}+ \nabla f^{(k+1)}, 
{1 \over \lambda}),
\nonumber \\
\mbox{\rm shrink}(x,\gamma) = {x \over |x|} 
\mbox{\rm max}(|x|-\gamma,0).
\nonumber
\end{eqnarray}

The filtered fields  reproduced in Fig. \ref{fig8} have been produced setting $f^{obs}
=\varepsilon_{xx}^{0,av}$ over the 2D grid in the plane XY, relabeling
to transform it into a 1D vector, and  using the algorithm:
\begin{itemize}
\item Initial guess $f^{(0)} = f^{obs}$, $d^{(0)}=0$, $b^{(0)}=0.$
\item While $\| f^{(k)} - f^{(k-1)} \|_{L_2} > Tol$
\begin{itemize}
\item $f^{(k+1)} = G^{(k)}$, where, for $j=1, \ldots, M,$
\begin{eqnarray}
 G^{(k)}_j = {\lambda \over \mu \!+\! 2\lambda}
\Big(f_{j+1}^{(k)} + f_{j-1}^{(k)} \!-\! (d_j^{(k)}\!\!-\!d_{j-1}^{(k)}) 
\!+\! (b_j^{(k)}\!\!-\!b_{j-1}^{(k)})\Big)  \!+\! {\mu \over \mu 
\!+\! 2 \lambda} f_j^{obs}, 
\nonumber
\end{eqnarray}
with  $\nabla f^{(k+1)}_j = f^{(k+1)}_{j+1} - f^{(k+1)}_j,$
\item $d^{(k+1)} = {\rm shrink}(b^{(k)} + \nabla f^{(k+1)}, {1\over
\lambda})$,
\item $b^{(k+1)} = b^{(k)} + \nabla f^{(k+1)} - d^{(k+1)}$.
\end{itemize}
\item If $\| f^{(k)} - f^{(k-1)} \|_{L_2} <Tol$, we set $f^{fil}=f^{(k)}.$
\end{itemize}

The resulting fields are smooth enough to be plugged in equations
(\ref{plategrowth1bis})-(\ref{plategrowth2bis}) through (\ref{strain})-(\ref{stress})
without causing numerical instability, 
allowing us to reproduce behaviors that resemble observed patterns.

\begin{figure}[!hbt]
\centering
 (a) \hskip 3cm (b) \\
\includegraphics[height=3cm]{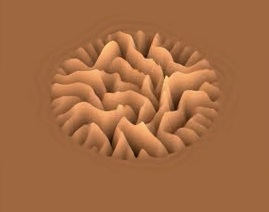}
\includegraphics[height=3cm]{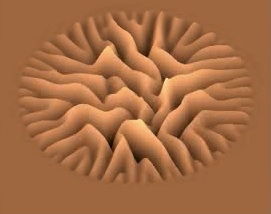}
\caption{In silico wrinkle coarsening and successive branching in a biofilm spreading on agar. The height of the wrinkles ranges from $-5$ dx to $5$ dx and the final radius is about $250$ dx, dx being the size of the tiles in the spatial discretization employed to evaluate the cellular activity.}
\label{fig9}
\end{figure}

Our simulations of biofilm behavior alternate steps in which we update the configuration of biofilm tiles,  creating and killing cells, and then evaluate the resulting stresses, 
with steps in which the biofilm shape is deformed as determined by the F\"oppl-Von Karman equations,  see \cite{carpiopre} for details.
Figure \ref{fig9} shows wrinkles coarsening and opening up in radial branches.
This phenomenon is associated to compression fronts expanding at certain speeds. Other observed arrangements, such as wrinkled coronas, that is, a corona of radial wrinkles surrounding a central core \cite{chaimrs}, can be reproduced varying the Young modulus as usual in corona instabilities: a swollen corona with diminished Young modulus around a harder core \cite{carpiopre}. Two phase models have been proposed to describe swelling processes in \cite{seminarapnas}.
However, an adequate way to handle water migration processes is still missing in our description. 

\section{Conclusions}
\label{sec:conclusions}

Summarizing, we have shown that inserting in plate and rod models  
information from the cellular activity and the interaction with the environment  
we gain insight on biofilm shapes. 

Coupling discrete rod models  to the forces exerted by surrounding fluids
and incorporating external constraints such as  the presence of walls and 
constrictions, we are able to reproduce experimentally observed filamentary
patterns.
Helical biofilms in flows arise as elastic instabilities that coarsen as the length 
of the biofilm thread increases due to biomass production and finally wrap 
around tube walls reducing their pitch to accommodate more biomass. The
main role of the flow in this case is to promote biofilm filament nucleation past 
constrictions and to provide mechanisms for twist at the filament edges.
Instead, biofilm threads are seen to cross 3D corner microflows to join opposite 
corners as a result of the interaction fluid/structure, in agreement with previous
2D studies.

Whereas biofilms in flows tend to form filaments, biofilms on agar surfaces often 
spread forming wrinkled patterns. Successive radial wrinkle branching is associated
to expanding compression fronts and can be reproduced by inserting residual stresses 
caused by cell division and death in  F\"oppl-Von Karman descriptions of the out of 
plane displacements of the interface biofilm/agar. These residual stresses are obtained 
from growth tensors, computed here using the information on the cellular activity provided 
by a simple cellular automata model. To avoid instability caused by stochasticity
and be able to visualize the spatial variations caused by the cellular activity, such 
residual stresses are smoothed out combining ensemble averages and denoising 
algorithms. More realistic models of  cellular activity could be considered at the
expense of increasing the computational cost.
Additional processes affecting biofilm shapes such as water migration 
through  the agar/biofilm system are missing in our description and should
be the object of further consideration. \vskip 2mm

{\bf Acknowledgement.}
This research has been supported by MINECO
grant No. MTM2014-56948-C2
and project C-ICT/3285 of the UE FP7.

\end{document}